\documentclass[aps, twocolumn,superscriptaddress,floatfix,preprintnumbers]{revtex4-1}
\usepackage{graphicx,amssymb}
\usepackage{soul}
\usepackage{xcolor}
\setstcolor{blue}
\usepackage{amsmath}
\usepackage{float}
\usepackage{amsmath,amssymb}
\usepackage{mathtools}
\usepackage{amssymb}
\usepackage{simplewick}
\usepackage{physics}
\usepackage{color}
\usepackage{bm}

\usepackage{hyperref}
\hypersetup{
	colorlinks=true,
	linkcolor=blue,
	filecolor=magenta,      
	urlcolor=cyan,
	citecolor=red,
}

\renewcommand{\S}{{\rm S}}
\newcommand{\N}{{\rm N}}
\newcommand{\I}{{\rm I}}

\renewcommand{\vec}[1]{\mathbf{#1}}

\renewcommand{\Re}{\mathop{\rm Re}}

\begin{document}
\title{Optimized  proximity thermometer for ultrasensitive detection: Role of an ohmic electromagnetic environment}
\author{Danilo Nikoli\'c}
\affiliation{Fachbereich Physik, Universit\"at Konstanz, D-78457 Konstanz, Germany}
\affiliation{Institut f\"ur Physik, Universit\"at Greifswald, Felix-Hausdorff-Strasse 6, 17489 Greifswald, Germany}
\author{Bayan Karimi}
\affiliation{QTF Centre of Excellence, Department of Physics, Faculty of Science, University of Helsinki, FI-00014 Helsinki, Finland}
\affiliation{ QTF Centre of Excellence, Department of Applied Physics, Aalto University School of Science, P.O. Box 13500, 00076 Aalto, Finland}
\author{Diego Subero Rengel}
\affiliation{ QTF Centre of Excellence, Department of Applied Physics, Aalto University School of Science, P.O. Box 13500, 00076 Aalto, Finland}
\author{Jukka P. Pekola}
\affiliation{ QTF Centre of Excellence, Department of Applied Physics, Aalto University School of Science, P.O. Box 13500, 00076 Aalto, Finland}
\author{Wolfgang Belzig}
\affiliation{Fachbereich Physik, Universit\"at Konstanz, D-78457 Konstanz, Germany}
\date{\today}

\begin{abstract}
We propose a mesoscopic thermometer for ultrasensitive detection based on the proximity effect in superconductor-normal metal (SN) heterostructures. The device is based on the zero-bias anomaly due to the inelastic Cooper pair tunneling in an SNIS junction (I stands for an insulator) coupled to an ohmic electromagnetic (EM) environment. The theoretical model is done in the framework of the quasiclassical Usadel Green's formalism and the dynamical Coulomb blockade. The usage of an ohmic EM environment makes the thermometer highly sensitive down to very low temperatures, $T \lesssim 5~$mK. Moreover, defined in this way, the thermometer is stable against small but nonvanishing voltage amplitudes typically used for measuring the zero-bias differential conductance in experiments. Finally, we propose a simplified view, based on an analytic treatment, which is in very good agreement with numerical results and can serve as a tool for the development, calibration, and optimization of such devices in future experiments in quantum calorimetry.
\end{abstract}

\maketitle
\section{Introduction}

Any temperature-dependent parameter, particularly one that varies monotonically, can provide a fundamental framework for thermometry~\cite{OVL, Quinn}. In the past decades, several techniques for measuring the local temperature at nanoscale have been developed~\cite{Aumentado1999,Jiang2003,Jiang2005,Mavalankar, Maradan, Libin, JP2, Nicola, Borzenets, Du}. In previous works~\cite{ZBA,Karimi2020}, we have proposed a technique based on the temperature-dependent proximity effect in a superconductor-normal metal-insulator-superconductor ($\S\N\I\S$) hybrid junction yielding sensitive thermometry with ultra-low dissipation. The thermometer was based on the zero-bias anomaly (ZBA) in the current-voltage ($I$-$V$) characteristics of such junctions. This on-chip technique has proven to be well-suited for detecting tiny heat currents through calorimetry, and also for performing fast thermometry towards the lowest temperatures in mesoscopic systems.

\begin{figure}[t!]
	\centering
	\includegraphics[width=\linewidth]{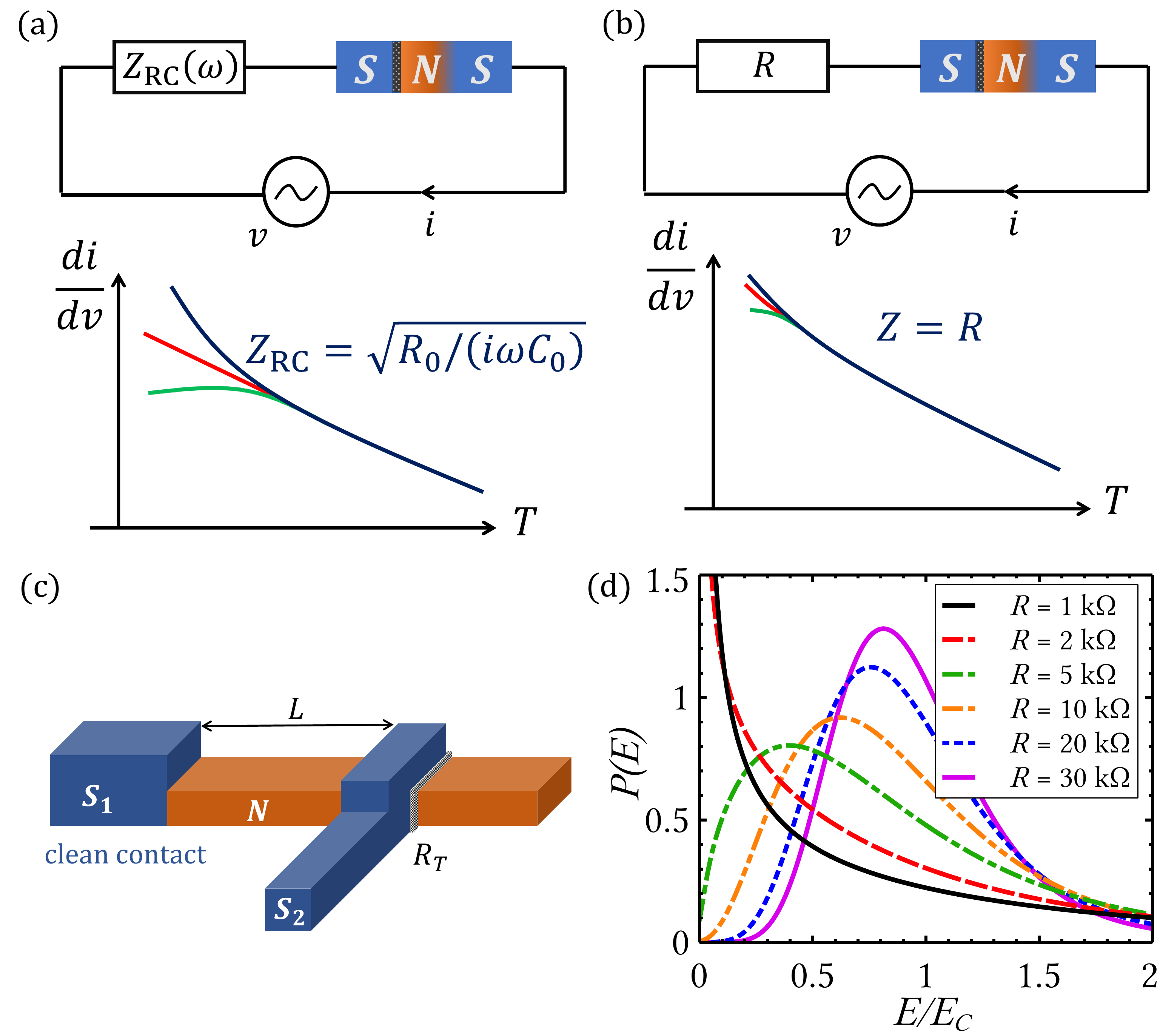}
	\caption{ (a) Schematic view of the circuit examined in Ref.~\cite{Karimi2020} where an SNIS proximity structure is coupled to an infinite $RC$ transmission line. Lower panel shows the low-bias conductance vs  temperature for different bias voltage amplitudes. (b) The corresponding circuit involving an ohmic environment instead. The low-bias conductance vs temperature is robust against nonzero voltage amplitudes typically unavoidable in experiments, making the proposed device stable and well-defined. (c) A scheme of the SNIS junction studied in the paper. (d) The $P(E)$ function for an ohmic electromagnetic environment shown in panel (b) for various resistances $R$. For certain resistances the zero-bias conductance does not lose sensitivity at low temperatures.}
	\label{fig:intro}
\end{figure}

The theoretical model presented in Ref.~\cite{Karimi2020} is based on the superconducting proximity effect and dynamical Coulomb blockade. The latter assumes the presence of an electromagnetic (EM) environment which was modeled as an infinite resistor-capacitor ($RC$) transmission line with effective (fitting) parameters [see the upper plot in Fig.~\ref{fig:intro}(a)]. Defined in this way, our model has quantitatively reproduced the experiments. However, we have found that the thermometer is not reliable at very low temperatures since it is highly sensitive to the  nonzero voltage amplitudes, $\delta V$, used for measuring the ZBA [see the lower plot in Fig.~\ref{fig:intro}(a)]. Namely, in calorimetric measurements the ZBA is typically measured at the small but nonvanishing lock-in voltage excitation, here referred to as $\delta V$. Hence, to have a well-defined ZBA thermometer, it is required to minimize the dependence of the ZBA on $\delta V$, and throughout the paper we say that the response is “robust” when the measured conductance does not depend strongly on $\delta V$. Besides this, the developed theory was phenomenological since the EM environment was not fabricated on purpose.

 In the present work, we propose a device based on an ohmic EM environment that overcomes the mentioned issues [see the upper plot in Fig.~\ref{fig:intro}(b)]. Namely, as we shall show the thermometer defined in this way is robust (in the sense discussed above) against nonzero voltage amplitudes~[see the lower plot in Fig.~\ref{fig:intro}(b)], especially in the experimentally relevant range $ \delta V\lesssim 0.5~\mu$V, and it is well defined since an ohmic impedance can be experimentally fabricated on purpose. 

The thermometer we propose here is based on an SNIS proximity junction schematically shown in Fig.~\ref{fig:intro}(c). A semi-infinite (total length is much larger than the superconducting coherence length; $\xi$) quasi-one-dimensional normal metal wire $\N$ (orange) is proximitized by a BCS superconductor ($\S_1$; blue) via clean contact. On distance $L$ from the clean contact another superconductor ($\S_2;$ blue) is put on top of the N wire via good tunnel contact of resistance $R_T$ and this structure acts as a thermometer. As already mentioned, the EM environment is modeled as an ohmic impedance [see Fig.~\ref{fig:intro}(b)]. As a result, we end up with a well-defined experimentally feasible thermometer. Namely, as we shall show in the subsequent sections, the ZBA, mediated by the inelastic tunneling of Cooper pairs, in such contacts scales monotonically with the temperature down to very low values, $T\lesssim 5$~mK. In addition, the device displays robustness against nonzero voltage amplitudes.

The paper is organized as follows. After the Introduction, in Sec.~\ref{sec:Theory} we provide the microscopic model of the proposed device based on the quasiclassical Green's function method and dynamical Coulomb blockade. In Sec.~\ref{sec:Discussion} we discuss the results of numerical calculations focusing on the operating regime of the thermometer, i.e., the ZBA vs temperature. In addition, here we provide a simplified analytic model that can be used as a calibration tool for the device. Finally, in Sec.~\ref{sec:Concludions} we enclose our discussion by giving concluding remarks and perspectives.  
\section{Theoretical framework}\label{sec:Theory}
As already mentioned above, to obtain a full model of the proposed device we need to account for two key ingredients: (i) the proximity effect in an SNIS junction described in the framework of the quasiclassical Green's formalism and (ii) the inelastic Cooper pair tunneling in the system caused by the dynamical Coulomb blockade. In this section, we discuss these. 
\subsection{Usadel formalism: SNIS proximity junction}\label{sec:Usadel}
The superconducting proximity effect is conveniently described by making use of the quasiclassical Green's function method~\cite{Eilenberger1968,Larkin1969, RammerSmith,Belzig1999}. In our model, all metallic parts are in the diffusive limit which assumes $\ell \ll \xi$, where $\ell=v_F \tau$ and $\xi=\sqrt{\hbar D/(2\Delta_0)}$ are the elastic mean free path and the superconducting coherence length, respectively.  For an aluminum superconductor, typical values for these are $\ell_\mathrm{Al} \sim 20$ nm and $\xi_\mathrm{Al} \sim 150$~nm. In equilibrium, the Green's function is a $2\times 2$ matrix in particle-hole space
\begin{equation}
\label{eqn:G_matrix}
	\hat{G}(\vec{r}) = \begin{pmatrix}
	G(\vec{r})& F(\vec{r})\\
	F^\dagger(\vec{r}) & -G(\vec{r})
	\end{pmatrix},
\end{equation}
 which is subject to the normalization $\hat{G}^2(\vec{r})=\hat{\tau}_0\implies G^2(\vec{r})+F(\vec{r})F^\dagger(\vec{r})=1$ and satisfies the Usadel equation~\cite{Usadel1970}
\begin{equation}
	\label{eqn:Usadel}
	\hbar D \bm{\nabla}\left[\hat{G}(\vec{r})\bm{\nabla}\hat{G}(\vec{r})\right] = \left[\omega_n \hat{\tau}_3 + \hat{\Delta}(\vec{r}), \hat{G}(\vec{r})\right].
\end{equation}
Here, $\vec{r}$ is the center-of-mass coordinate, $\omega_n=(2n+1)\pi k_BT_\mathrm{J}$ are fermionic Matsubara frequencies with $n=0,\pm 1, \pm 2, \dots$ and the electron temperature $T_\mathrm{J}$, $D=v_F\ell/3$ is the diffusion coefficient of the material, and $[\cdot~,~\cdot]$ denotes a commutator. In addition, $\hat{\tau}_i$ are the Pauli matrices in particle-hole space and $\hat{\Delta}$ is the gap matrix given by
\begin{equation}
	\hat{\Delta}(\vec{r}) = \begin{pmatrix}
	0 & \Delta \\
	\Delta^* & 0
	\end{pmatrix} = 
	\Delta(\vec{r})\begin{pmatrix}
	0 & e^{i\phi(\vec{r})}\\
	e^{-i\phi(\vec{r})} & 0
	\end{pmatrix},
\end{equation}
with $\phi(\vec{r})$ being the superconducting phase. 

Once known, the Green's function provides us with the full information about the equilibrium properties of the system under study, e.g., the supercurrent. If we deal with the system consisting of  two superconductors separated by an insulating barrier of resistance $R_T$, the supercurrent is given by the standard sinusoidal current-phase relation with the critical current~\cite{AmbegaokarBaratoff,Werhamer1966,Larkin1967,Belzig2001}
\begin{equation}
	\label{eqn:Ic_general}
	I_c = \frac{2\pi k_BT_\mathrm{J}}{eR_T}\sum_{n\geq0} F_1(\omega_n)F_2(\omega_n),
\end{equation}
where the index 1 (2) refers to the left (right) superconducting electrode. This formula will be used for calculating the Josephson energy of the thermometer, $E_J=\hbar I_c/(2e)$. Finally, we stress that the gap here is not calculated self-consistently and for its temperature dependence we use the interpolation formula $\Delta(T_\mathrm{J}) = \Delta_0 \tanh(1.74\sqrt{T_c/T_\mathrm{J}-1})$, where $\Delta_0$ denotes the gap at $T_\mathrm{J}=0$ and $T_c=\Delta_0/(1.764k_B)$ is the superconducting transition temperature~\cite{Gap}. As we show below, Eq.~\eqref{eqn:Ic_general} applies to the case of an SNIS system sketched in Fig.~\ref{fig:intro}(c). Namely, the proximitized N wire behaves as a superconductor inheriting some superconducting properties including the phase from the $\S_1$ lead.

As anticipated, the proposed device is based on an SNIS junction sketched in Fig.~\ref{fig:intro}(c), where a semi-infinite $\S_1\N$ proximity contact is coupled to another BCS superconductor $\S_2$ on distance $L$. The $\N\S_2$ interface has high resistance $R_T=1/\mathcal{G}_T$ and therefore Eq.~\eqref{eqn:Ic_general} applies. To describe the proximity effect in N induced by $\S_1$, we employ the so-called $\theta$-parametrization, where $G=\cos\theta$ and $F=\sin\theta$, and Eq.~\eqref{eqn:Usadel} adopts the form~\cite{Belzig1999,Nazarov1996}
\begin{equation}
	\label{eqn:Usadel_theta}
	\frac{\hbar D}{2} \dv[2]{\theta_n(x)}{x} = \omega_n\sin \theta_n(x) - \Delta(x)\cos \theta_n(x),
\end{equation}
 Here we neglect the inverse proximity effect, i.e., for the gap we take a stepwise potential, $\Delta(x)=\Delta\Theta(-x)$, where $\Theta(x)$ stands for the Heaviside step function. For simplicity, we assume the diffusion coefficient to be the same in both materials, $D_S=D_N=D$.  To obtain a full solution to the problem, the Usadel equation should be supplemented by the appropriate boundary conditions, $\theta_n(x=0-) = \theta_n (x=0+)$ and $\sigma_S\partial_x\theta_n|_{x=0-} = \sigma_N\partial_x\theta_n|_{x=0+}$, where $\sigma_{S/N}$ is the normal-state conductivity of the S/N region~\cite{KuprianovLukichev}.  Two other boundary conditions read $\theta_n(x\to \infty) = 0$ and $\theta_n(x\to -\infty) = \theta_n^S$, with  $\theta_n^S=\atan(\Delta/\omega_n)$  being the BCS bulk solution. 
 
 Under the assumptions given above, the solution of Eq.~\eqref{eqn:Usadel_theta} in the N wire reads~\cite{Zaikin1981,Belzig1996}
\begin{equation}
	\label{eqn:theta_SN_inf}
	\theta_n(x) = 4\atan\left[\tan(\frac{\theta_n^0}{4})\exp[-\sqrt{2\omega_n/(\hbar D)}x]\right], 
\end{equation}
where the $\theta_n^0$ function satisfies the equation
\begin{equation}
	\label{eqn:BC}
	\sin\left(\frac{\theta_n^0-\theta_n^S}{2}\right)= -\gamma\left(\frac{\omega_n^2}{\omega_n^2+\Delta^2}\right)^{1/4}\sin\left(\frac{\theta_n^0}{2}\right).
\end{equation}
Here, $\gamma=\sigma_N\sqrt{D_S}/(\sigma_S\sqrt{D_N})$ is the so-called diffusivity mismatch parameter, which in our case measures the mismatch in the normal-state conductivities of two materials since we assume $D_S=D_N=D$. 

\begin{figure}[t!]
	\centering
	\includegraphics[width=7.2cm]{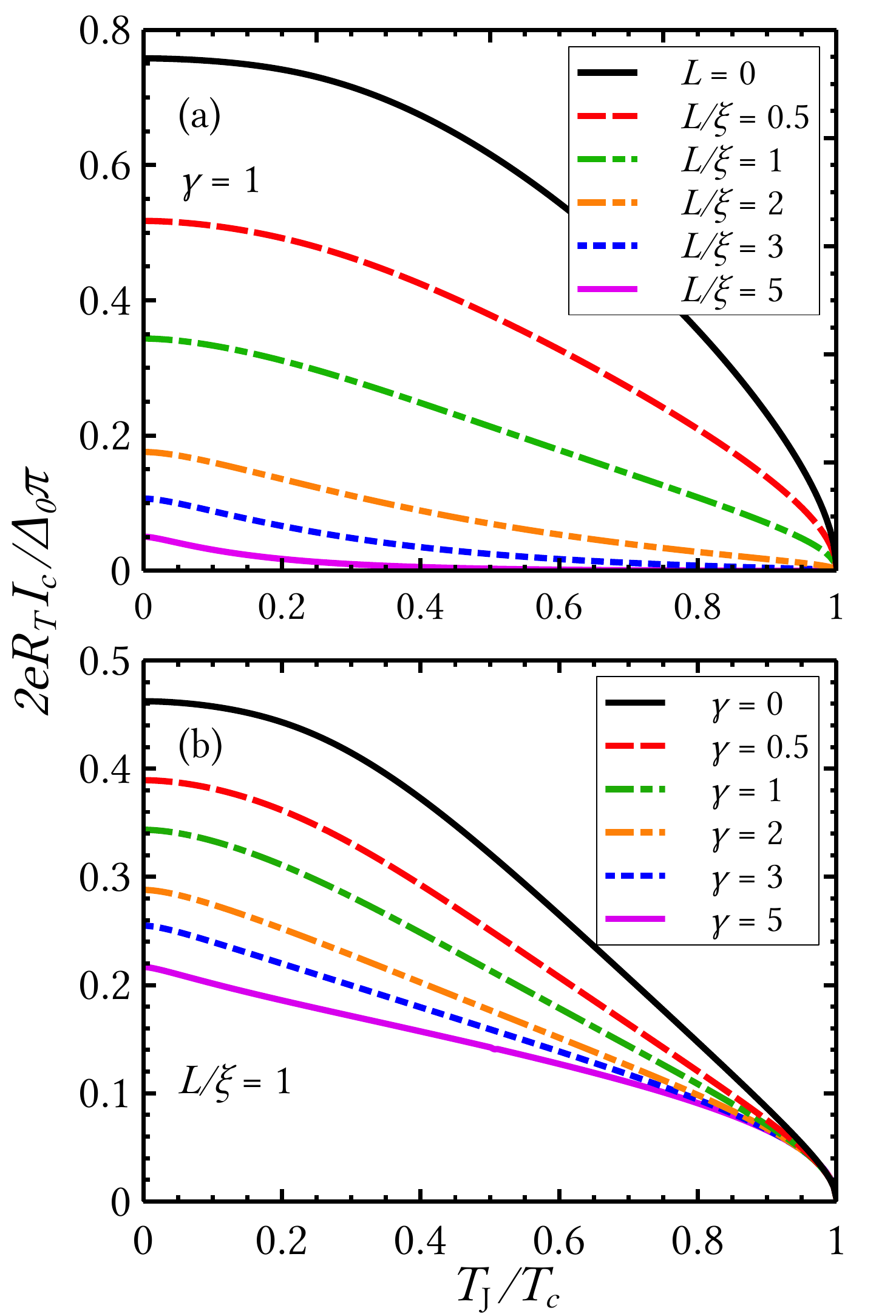}
	\caption{The critical current $I_c$ in an SNIS thermometer as a function of the junction's temperature $T_\mathrm{J}$ for (a) various distances $L$ between the clean and the tunnel contact  and $\gamma=1$ and (b) various $\gamma$ and $L=\xi$.}
	\label{fig:Ic}
\end{figure}

Having obtained the solution in the N wire, we can calculate the critical current across the $\N\S_2$ interface by applying Eq.~\eqref{eqn:Ic_general}, i.e., 
\begin{equation}
	\label{eqn:Ic_SNIS}
	I_c(L)= \frac{2\pi k_B T_\mathrm{J}}{eR_T} \sum_{n\geq 0} \frac{\Delta\sin\theta_n(x=L)}{\sqrt{\omega_n^2+\Delta^2}},
\end{equation}
where $L$ is the distance between the superconductors $\S_1$ and $\S_2$ [see Fig. 1(c)]. Note that we neglect the inverse proximity effect in $\S_2$, i.e., we assume it as a bulk homogeneous BCS superconductor, $F_2 = \Delta/\Omega_n$, where $\Omega_n=\sqrt{\omega_n^2+\Delta^2}$.  The resulting critical current is presented in Fig.~\ref{fig:Ic}. Panel (a) shows the $I_c(T_\mathrm{J})$ function for various distances between the superconductors $L$ and $\gamma=1$. The critical current is getting suppressed on longer distances (see the solid violet line) since the proximity effect weakens as the Cooper pairs propagate deeper into the N wire from the $\S_1\N$ interface. Panel (b) shows the same quantity for $L=\xi$ and different values of the $\gamma$ parameter. Increasing $\gamma$ significantly suppresses the critical current. 
\subsection{P(E) theory: Ohmic EM environment}
\label{sec:EM}
The second ingredient necessary for calculating the tunneling current in the SNIS thermometer is information about an EM environment coupled to the junction. A Josephson junction of sufficiently high charging energy, $E_C=2e^2 /C\gg E_J$, $C$ being the junction capacitance, can feature the single Cooper pair tunneling carrying the current~\cite{Averin1990,IngoldNazarov1992}
\begin{equation}
	\label{eqn:I_s}
	I_s(V) = \frac{\pi e E_J^2}{\hbar} \left[P(2eV)-P(-2eV)\right].
\end{equation}
Here, $E_J=\hbar I_c/(2e)$ is the Josephson energy and $P(E)$ is the probability for a Cooper pair to emit a photon of energy $E$ to the EM environment during the inelastic tunneling across the junction. Note that the Josephson energy directly depends on the junction's properties through Eq.~\eqref{eqn:Ic_SNIS} [see also Eq.~\eqref{eqn:Ic_general}]. The $P(E)$ function is given by
\begin{equation}
	\label{eqn:P(E)}
	P(E)=\frac{1}{2\pi\hbar}\int_{-\infty}^{\infty}dt \exp\left[4J(t)+\frac{i}{\hbar}Et\right],
\end{equation}
where $J(t)=\langle[{\varphi}(t)-{\varphi}(0)]{\varphi}(0)\rangle$
is the equilibrium correlation function of the phase $\varphi(t)=(e/\hbar)\int_{-\infty}^t V(t')dt'$ with $V(t)$ being the voltage across the junction. The $J(t)$ function depends on the total impedance of the system seen by the junction, $Z_t(\omega)$, and reads
\begin{align}
	\label{eqn:J(t)}
	J(t)=\, &2\int_0^\infty \frac{d\omega}{\omega}\frac{\Re[Z_t(\omega)]}{R_K}\\&\times
	\bigg\{\coth\bigg(\frac{\hbar\omega}{2k_{B}T_\mathrm{EM}}\bigg)[\cos(\omega t)-1]-i\sin(\omega t)\bigg\}\nonumber,
\end{align}
where $R_K=h/e^2\approx 25.8$ k$\Omega$ is the von Klitzing constant denoting the resistance quantum and $T_\mathrm{EM}$ is the environmental temperature. Finally, the total impedance of the system $Z_t(\omega)$ reads
\begin{equation}
	\label{eqn:totalZ}
	Z_t(\omega)=\frac{1}{i\omega C+Z^{-1}(\omega)},
\end{equation}
where $C$, as introduced earlier, is the capacitance of the junction and $Z(\omega)$ is the impedance of the EM environment itself. As already mentioned, the usage of an infinite $RC$ transmission line succeeds in describing a proximity thermometer and quantitatively reproduces experimental data~\cite{Karimi2020}. However, this approach is phenomenological, since no EM environment has been fabricated on purpose in the experiment.

In the present model, the EM environment is assumed to be an ohmic impedance of resistance $R$, where
\begin{equation}
		\label{eqn:EM_impedance}
		\frac{\Re[Z_t(\omega)]}{R_K}=\frac{1}{R_K}\frac{R}{1+(\omega RC)^2}=\frac{\rho}{1+(\omega/\omega_R)^2},
\end{equation}
with $\rho=R/R_K$ and
\begin{equation}
	\omega_R=\frac{1}{RC}=\frac{1}{4\pi\rho}\frac{E_C}{\hbar}.
\end{equation}
The corresponding $P(E)$ function for several environmental resistances $R$ and temperature $k_BT_\mathrm{EM}/E_C=0.01$ is shown in Fig.~\ref{fig:intro}(d)~\cite{IngoldNazarov1992}.  As one may notice low resistances $(R< R_K/8\approx 3.23~\mathrm{k}\Omega)$ lead to a divergence of the function at low energies (see the solid black and dashed red lines). According to Eq.~\eqref{eqn:I_s} this would lead to a divergence of the current itself, which means that the latter formula is out of range of validity. Namely, it can be shown that the formula for the tunneling current given in Eq.~\eqref{eqn:I_s}, as a perturbative result in $E_J$, breaks down for sufficiently small environmental resistances, $R<R_K/8$, and for low energies~\cite{IngoldNazarov1992, Grabert1998}. In this case, one is required to calculate the contributions from all orders  in $E_J$ as it was done in Ref.~\cite{Grabert1999}. However, in the regime $R>R_K/8$, Eq.~\eqref{eqn:I_s} is valid and it can be used for describing the Cooper-pair-mediated ZBA in $I$-$V$ responses.

\section{Description of the thermometer}\label{sec:Discussion}
The following discussion is mainly based on Eq.~\eqref{eqn:I_s}. It is important to stress that both the junction and the environmental contribution depend on temperatures that do not need to be equal in general. As already indicated in the previous section, we term these $T_\mathrm{J}$ and $T_\mathrm{EM}$, respectively. Note that the natural energy units for these are the superconducting gap, $\Delta_0$, and the charging energy of the junction, $E_C$, respectively. 
All results are obtained for $E_C=\Delta_0$, where we take the zero-temperature gap value for aluminum, $\Delta_0\approx 200~ \mu eV$, which corresponds to the critical temperature $T_c\approx 1.3$ K. In all calculations $E_J<0.1 E_C$, thus, we are in the range of validity of Eq.~\eqref{eqn:I_s}.
\subsection{I-V curves: ZBA}
\begin{figure}[t!]
	\centering
	\includegraphics[width=7.5cm]{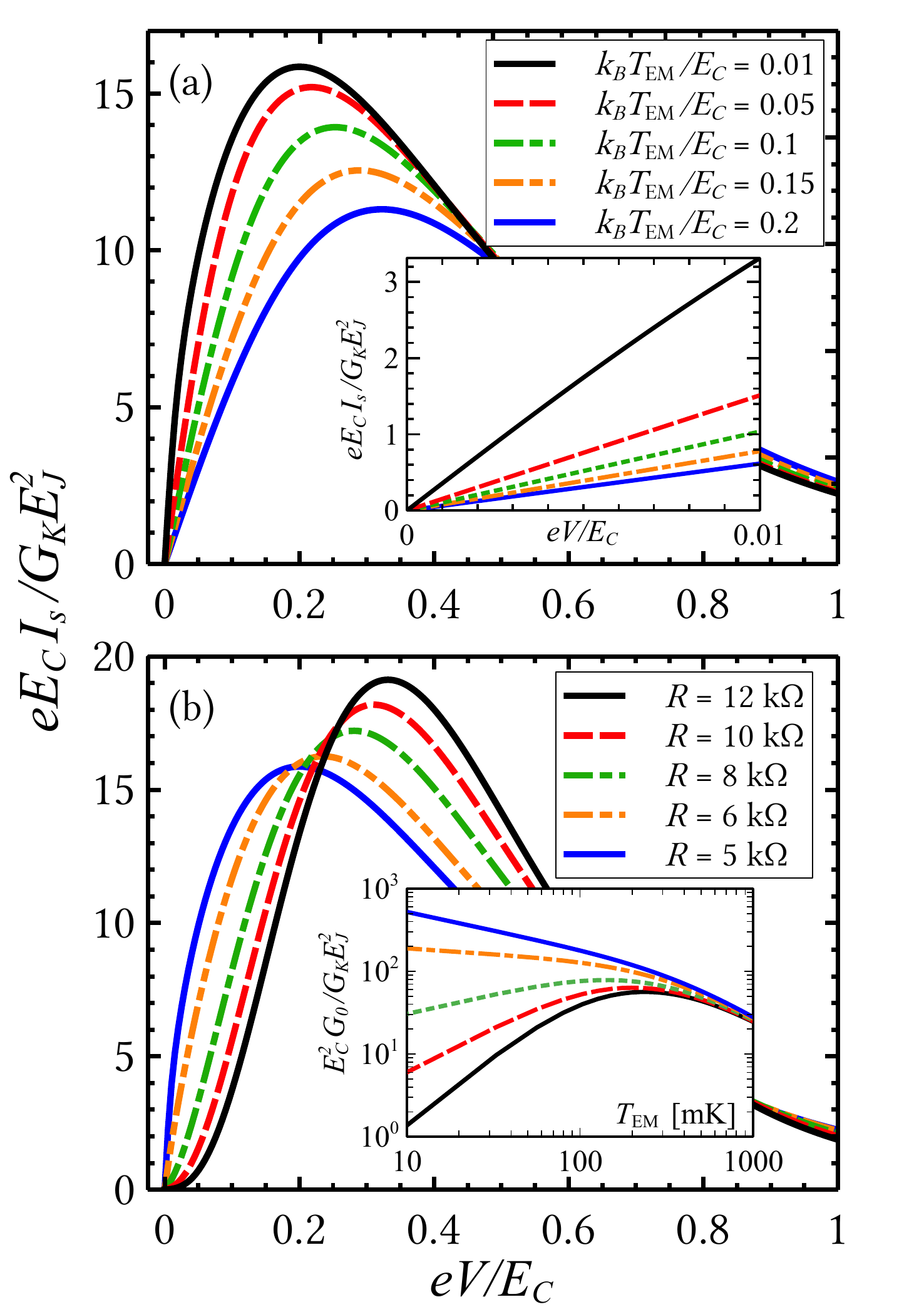}
	\caption{(a) The $I$-$V$ characteristics of a junction of Josephson energy $E_J$ and charging energy $E_C$ coupled to an ohmic electromagnetic environment at various temperatures and $R=5$~k$\Omega$. The inset shows a closer look into $I$-$V'$s that are linear at low voltages. (b) The same quantity calculated for various environmental resistances $R$ and temperature $k_BT_\mathrm{EM}/E_C=0.01$. The inset shows the corresponding zero-bias conductance vs temperature.}
	\label{fig:IVs}
\end{figure}

Let us here for a moment ignore the properties of the junction itself and analyze only the $P(E)$ contribution to Eq.~\eqref{eqn:I_s}. The $I$-$V$ curves are shown in Fig.~\ref{fig:IVs}. Panel (a) shows the $I$-$V$ responses for various environmental temperatures $T_\mathrm{EM}$ and the resistance $R=5 $~k$\Omega$. Note that the current itself is scaled in units of $G_KE_J^2/(eE_C)$, where $G_K=e^2/h=1/R_K$ is the conductance quantum and $E_J$ is the Josephson energy which we do not calculate here. We just assume that the temperature of the junction, $T_\mathrm{J}$, is fixed and in general different than the environmental one, $T_\mathrm{EM}$. We see that $I$-$V'$s are getting suppressed with increasing $T_\mathrm{EM}$. The inset shows a closer look into $I$-$V'$s that are linear for small voltages. This important feature indicates the robustness of the low-bias differential conductance against nonzero bias voltages, the effect we discuss explicitly below. 

Panel (b) shows the same function for various environmental resistances $R$ and temperature $k_BT_\mathrm{EM}/E_C=0.01$. The maximum is shifted towards higher voltages, i.e., for larger $R$ the dynamical Coulomb blockade is enhanced. The corresponding zero-bias differential conductance $G_0$ vs temperature $T_\mathrm{EM}$ is shown in the inset. Apparently, lower resistances lead to monotonically decreasing behavior (see the solid blue and dash-dotted orange lines) making our model a suitable thermometer. Furthermore, the responsivity of the thermometer, $\mathcal{R}=\abs{dG_0/dT_\mathrm{EM}}$, is enhanced with decreasing resistance. However, we emphasize that $R$ cannot be arbitrarily small since the theory breaks down for $R<R_K/8 \approx 3.23$~k$\Omega$, as already discussed in Sec.~\ref{sec:EM}. Therefore, we expect the thermometer to display the best performances for resistances $R\sim 4-6~$k$\Omega$. Larger resistances, $R>6~$k$\Omega$, however, lead to a nonmonotonic behavior of the device (see, for instance, the black solid line) at certain temperatures; in this case $T\lesssim 200$ mK.

\subsection{Operating regime of the thermometer}

\begin{figure}[t!]
	\centering
	\includegraphics[width=\linewidth]{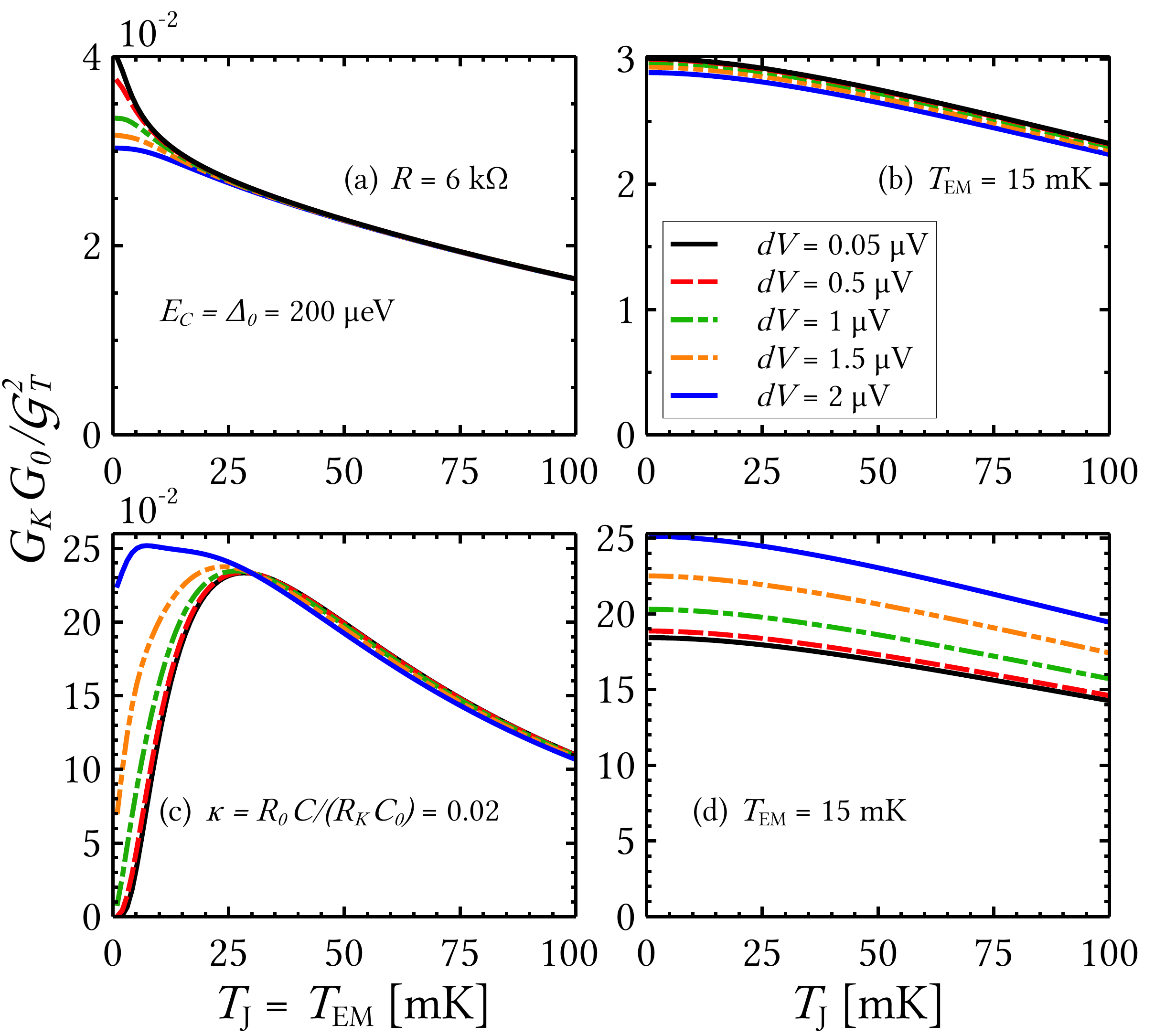}
	\caption{(color online) Effect of a nonzero voltage amplitude: The low-bias conductance $G_0$ as a function of temperature of the junction $T_\mathrm{J}$ in an SNIS junction coupled to an ohmic environment of resistance $R=6~$k$\Omega$ for (a) equal temperatures of the junction and the environment, $T_\mathrm{J}=T_\mathrm{EM}$ and (b) fixed environmental temperature $T_\mathrm{EM} = 15$ mK. Lower panels show the corresponding quantity in the case of an infinite $RC$ transmission line characterized by $\kappa=0.02$  (for more details, see Ref.~\cite{Karimi2020}). In all panels, the junction's parameters are $L=3\xi$ and $\gamma=1$. Note that the legend of panel (b) applies to all panels.}
	\label{fig:ZBA}
\end{figure}
The main question that arises in our discussion is how to optimize the proposed device. There are several criteria to do so; however, here we shall focus on two: (i) sensitivity of the proposed device at low temperatures and (ii) robustness against small but nonvanishing voltages whose presence is unavoidable in calorimetric measurements. Considering the temperature dependence, in what follows we distinguish two cases: (i)  the temperature of the junction $T_\mathrm{J}$ equals the temperature of the EM environment $T_\mathrm{EM}$ and (ii) the latter is fixed and heating processes change only the former. Under certain circumstances, the second scenario is experimentally relevant. 

 Figures~\ref{fig:ZBA}(a) and~\ref{fig:ZBA}(b) show the low-bias differential conductance as a function of temperature for different amplitudes of the applied voltages. The SNIS junction of length $L=3\xi$ and $\gamma=1$ is coupled to the ohmic EM environment of resistance $R=6~$k$\Omega$. Panel (a) assumes the same temperatures of the junction and the environment, $T_\mathrm{J}=T_\mathrm{EM}$. First, we notice that the thermometer does not lose sensitivity even at  very low temperatures, $T < 5~$mK, even for nonzero voltage amplitudes $\delta V$ [see the black solid and red dashed line in Fig.~\ref{fig:ZBA}(a)]. The second important feature is that the device is robust against $\delta V$ in a wide range of values, $\delta V\sim 0-2~\mu$eV, at reasonably low temperatures, $T\sim 10$ mK. This is especially significant in the experimentally most relevant regime, $\delta V\lesssim0.5~\mu$V, where the robustness is high even at $T < 5$ mK. In panel (b) we show the same quantity as a function of the junction's temperature $T_\mathrm{J}$ while the environmental temperature is fixed at $T_\mathrm{EM}=15~$mK. As one may notice, this scenario also supports the results presented in panel (a). However, in this regime, the thermometer loses sensitivity but at low temperatures, $T\sim 5-10~$mK.

For comparison, in Figs.~\ref{fig:ZBA}(c) and~\ref{fig:ZBA}(d) we present the characterization of the thermometer based on an infinite $RC$ transmission line described by $\kappa=R_0C/(R_KC_0)=0.02$, where $R_0$ and $C_0$ denote, respectively, the resistance and the capacitance per unit length of the line. Correspondingly, panel (c) shows the case of the equal temperatures of the junction and the environment, $T_\mathrm{J}=T_\mathrm{EM}$, while panel (d) assumes the fixed temperature of the environment, $T_\mathrm{EM}=15~$mK. Other parameters are the same as in panels (a) and (b). Such a device was examined in detail in Ref.~\cite{Karimi2020}.  Although displaying higher responsivity, a thermometer defined in this way is disadvantageous in two ways. First, it shows nonmonotonic behavior at significantly larger temperatures than the ohmic case discussed above; here  $T\lesssim 30~$mK [the solid black line in Fig.~\ref{fig:ZBA}(c)]. Second, it is sensitive to nonzero amplitudes at $T\lesssim 30~$mK. However, above a certain temperature threshold, in this case $T\sim 30~$mK, an $RC$ transmission line can be utilized as an EM environment and this was the regime considered in Ref.~\cite{Karimi2020}. On the other hand, if the environmental temperature is fixed, the thermometer loses sensitivity at very low temperatures, $T\sim 5-10~$mK, similarly to the ohmic case. However, the sensitivity to a nonzero voltage amplitude is still an issue [see Fig.~\ref{fig:ZBA}(d)].

\subsection{Calibration of the thermometer: Analytic formula}
Under certain assumptions, the picture presented above can be simplified and expressed in an analytic manner. The following approach can be utilized as an optimization and calibration tool in future experiments. 

 As reported in Ref.~\cite{Ingold1994}, the zero-bias conductance in the case of an ohmic EM environment at low temperatures can be expressed as
\begin{equation}
	\label{eqn:G0_approx}
	\frac{G_0}{G_K} = 4\pi^4\left(\frac{E_J}{E_C}\right)^2(4\rho)^{8\rho}\frac{\Gamma(4\rho)^2}{\Gamma(8\rho)}e^{-8\zeta\rho}\left(\frac{\beta_\mathrm{EM} E_C}{2\pi^2}\right)^{2-8\rho},
\end{equation}
where $\Gamma(x)$ denotes the Euler gamma function, whilst $\zeta=0.5772\dots$ is Euler's constant (conveniently denoted as $\gamma$, but we use different notation to avoid ambiguity with the mismatch parameter $\gamma$ used throughout the paper). In other words, the temperature dependence that comes from the environment only scales as $\sim T_\mathrm{EM}^{8\rho-2}$. Note that the formula above does not specify anything about the Josephson energy $E_J$.

To describe the proximitized N wire, we linearize the Usadel equation [see Eq.~\eqref{eqn:Usadel_theta}] in $\theta_n$. Namely, by supposing that $\theta_n$ is substantially suppressed in the normal metal, $\theta_n\approx\delta\theta_n\ll \theta_n^S$, and being close to the BCS bulk value in the superconducting electrode, $\theta_n \approx \theta_n^S -\delta\theta_n$, the Usadel equation \eqref{eqn:Usadel_theta} can be linearized in $\delta\theta_n$ and  solved analytically. Having obtained the solution of the proximity angle in N, we can calculate the critical current in the thermometer [see Eq.~\eqref{eqn:Ic_general}] arriving at the following equation for the Josephson energy:
\begin{equation}
	\label{eqn:Ej_linearized}
	E_J = \frac{h k_BT_\mathrm{J}}{2e^2R_T} \sum_{n\geq0}\frac{\Delta\sin\delta\theta_n}{\sqrt{\omega_n^2+\Delta^2}} ,
\end{equation}
where
\begin{equation}
\label{eqn:theta_approx}
	\delta\theta_n(x) = \frac{\theta^S_n\exp[-\sqrt{2\omega_n/(\hbar D)}x]}{1+\gamma\sqrt{\omega_n/\Omega_n}}
\end{equation}
with the corresponding notation as in Sec.~\ref{sec:Theory}. Note that strictly speaking the equations above are only valid in the case of small $\delta \theta_n(x)$, which is achieved for, e.g.,  $L\gg\xi$ and/or $\gamma\gg 1$. 
\begin{figure}[t!]
	\centering
	\includegraphics[width=7.5cm]{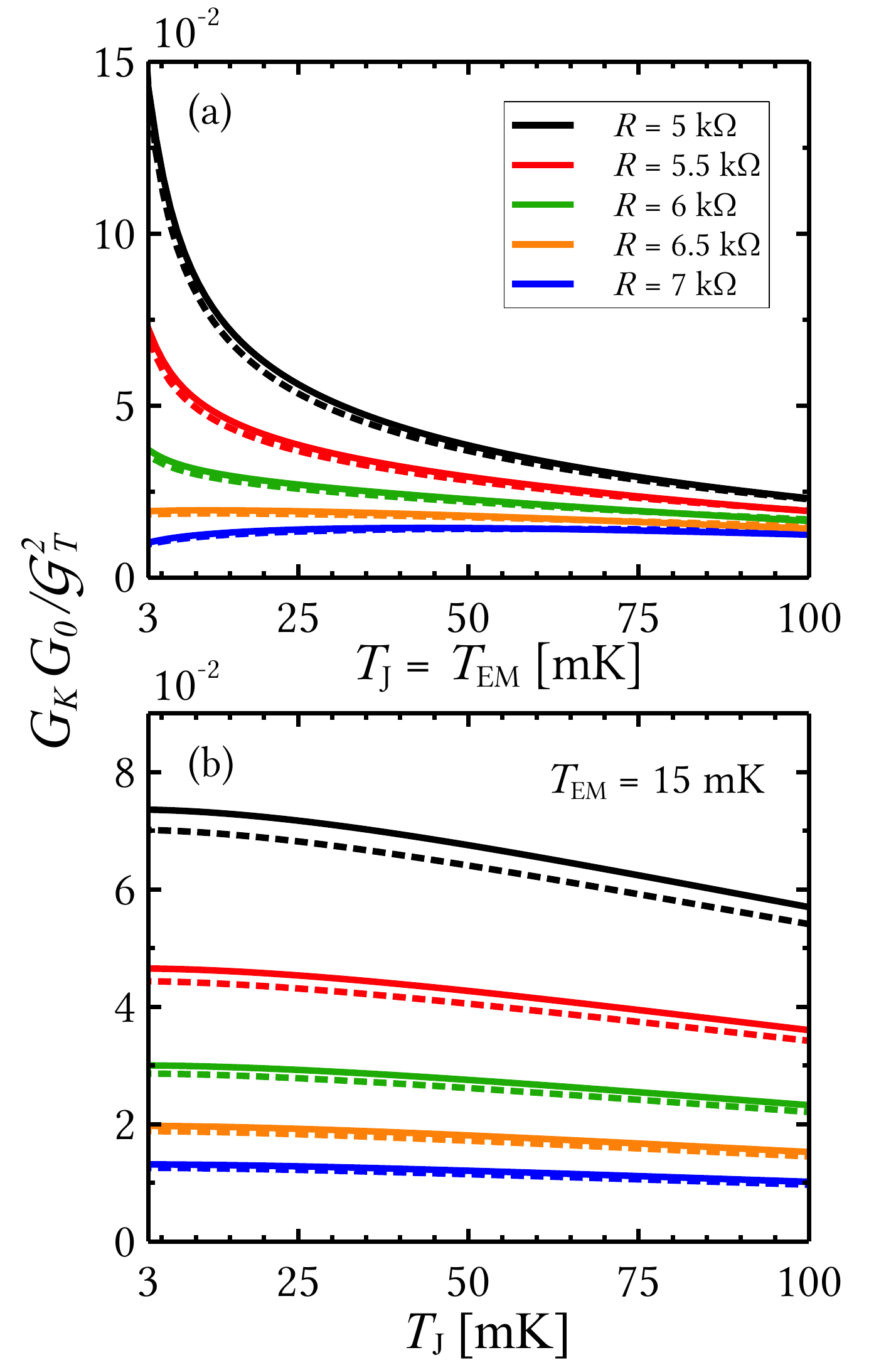}
	\caption{Calibration of the thermometer: The zero-bias conductance $G_0$ as a function of the junction's temperature $T_\mathrm{J}$ for various environmental resistances $R$ and (a) equal temperatures of the junction and the environment $T_\mathrm{J}=T_\mathrm{EM}$ and (b) fixed environmental temperature $T_\mathrm{EM}=15~$mK. Dotted lines correspond to the analytic formula obtained by combining Eqs.~\eqref{eqn:G0_approx}-\eqref{eqn:theta_approx}. Junction's parameters are $\gamma=1$ and $L=3\xi$.}
	\label{fig:Calibration}
\end{figure}

In accordance with the discussion from the previous sections, here we discuss two cases regarding the temperatures of the junction and the environment. In this respect, Fig.~\ref{fig:Calibration}(a) shows the zero-bias conductance $G_0$ as a function of temperature which is the same for the junction and the environment, $T_\mathrm{J}=T_\mathrm{EM}$. Different curves correspond to different environmental resistances $R$ and the junction's parameters are $L=3\xi$ and $\gamma=1$. The dotted lines represent the approximate result obtained by combining Eqs.~\eqref{eqn:G0_approx}-\eqref{eqn:theta_approx}. Apparently, our simplified model quantitatively reproduces the full theory with small deviations. Correspondingly, Fig.~\ref{fig:Calibration}(b) shows the same quantity as a function of the junction's temperature $T_\mathrm{J}$ while the environmental one is fixed at $T_\mathrm{EM}=15$ mK. Similarly to panel (a), the analytic formulas resemble the full theory [see the dotted lines in Fig.~\ref{fig:Calibration}(b)]. 
\section{Conclusions}\label{sec:Concludions}

We have proposed a proximity thermometer for ultrasensitive detection. The device is based on the Cooper-pair-mediated ZBA in an SNIS system coupled to an ohmic EM environment.  The calculations have been carried out in the framework of the quasiclassical Usadel Green's formalism and dynamical Coulomb blockade [the $P(E)$ theory].

Owing to the ohmic EM environment, the thermometer shows high responsivity and does not lose sensitivity even at low temperatures, $T\lesssim 5~$mK.  In addition, we have considered two distinct cases concerning the temperature: (i) the temperature of the junction equal to the environmental temperature, $T_\mathrm{J}=T_\mathrm{EM}$ and (ii) the heating effects only affect the junction's temperature $T_\mathrm{J}$ whereas the environmental one is fixed, $T_\mathrm{EM}=$ const.

Since calorimetric measurements are typically performed by applying small but nonvanishing voltage amplitudes, we have addressed this question explicitly. In this respect, our device shows robustness against nonzero voltages even at very low temperatures, especially in the experimental relevant range, $\delta V\lesssim 0.5~\mu$V, and therefore it can be utilized as a thermometer with well-defined characteristics.

Finally, in order to calibrate our device, we have provided a simplified analytic view based on the linearized Usadel equation which is in good agreement with the full theory. This approach can be easily implemented as a calibration and optimization tool in future experiments in quantum calorimetry~\cite{Avila2019,Karimi2020b,Kokkoniemi2020,Lee2020,Gumus2023}.

\acknowledgements

This work received funding from the European Union’s Horizon 2020 research and innovation programme under Marie Sklodowska-Curie actions (Grant No. 766025, QuESTech) and Deutsche Forschungsgemeinschaft (DFG; German Research Foundation) via SFB 1432 (Project No. 425217212).

\end{document}